\documentclass[prb,floatfix,twocolumn]{revtex4}
\usepackage[T1]{fontenc}
\usepackage[latin1]{inputenc}
\usepackage{graphicx} % Include figure files
\usepackage{dcolumn} %Align table columns on decimal point
\usepackage{bm} %bold math
\usepackage{ae,aecompl}
\usepackage{color}
  
\begin{document}
 
\title{Negative oxygen vacancies in HfO$_2$ as charge traps in high-k stacks} 
 
\author{J. L. Gavartin, D. Mu\~{n}oz Ramo, A. L. Shluger}
\affiliation{Department of Physics and
Astronomy, University College London, Gower Street, London WC1E 6BT,
UK} 
\author{G. Bersuker and B. H. Lee}
\affiliation{SEMATECH, 2706 Metropolis Dr., Austin, TX 78741, USA}
\date{\today}

\begin{abstract}
We calculated the optical excitation and thermal ionization energies of oxygen 
vacancies in m-HfO$_2$ using atomic basis sets, a non-local density functional 
and periodic supercell. The thermal ionization 
energies of negatively charged V$^-$ and V$^{2-}$ centres are consistent with 
values obtained by the electrical measurements. 
The results suggest that negative oxygen 
vacancies are the likely candidates for intrinsic electron traps in 
the hafnum-based gate stack devices. 
\end{abstract}

\maketitle
   
Hafnium based oxides are currently considered as a practical 
solution satisfying stringent criteria for integration of high-k materials in 
the devices in future technology nodes. However, high-k transistor 
performance is often affected by high and 
unstable threshold potential \cite{Young_1_2004}, $V_t$, and low carrier 
mobility 
\cite{Carter_aps2003}. These effects are usually attributed to a high 
concentration of charge traps and scattering centers in the bulk of the 
dielectric and/or at its interface with the silicon channel. 
Although the reported trap densities vary greatly with fabrication techniques, 
the majority of data point to existence of a specific intrinsic 
shallow electron centre common to all HfO$_2$ based stacks while some 
extrinsic defects, such as Zr substitution in HfO$_2$, have also been considered 
\cite{Bersuker_Nato_2006}. 

Oxygen vacancies are dominating intrinsic defects in the {\it bulk} of many 
transition metal oxides including HfO$_2$ and ZrO$_2$, and are thought to be 
also present in high concentrations in thin films. However, in spite of numerous 
experimental studies, evidence relating oxygen vacancies to measured 
characteristics of interface traps in high-k stacks is still mostly 
circumstantial. Therefore, accurate theoretical characterization of these
defects is highly desirable.  

Previous theoretical calculations of oxygen vacancy in HfO$_2$ and ZrO$_2$ 
reported 
the ground state properties obtained within local or semilocal approximations
to density functional theory (DFT) methods 
(see \cite{Robertson_repproghys2006,nano_book_2003} for a review). This 
approach, however, significantly underestimates band gaps, which hampers 
determining energies of defect levels with respect to the band edges and 
precludes identifying shallow defect states 
\cite{nano_book_2003,gavartin_jap2005,VandeWalle_jap2004}. As a result, most of 
the early local DFT calculations (except, perhaps, ref.\cite{Shen_IEDM2004}) 
failed to predict unambiguosly negative charge states of oxygen vacancy in 
HfO$_2$. Significant improvement was achieved by 
Robertson {\em et al.} 
\cite{Xiong_Robertson_MicEng_2005,Xiong_Robertson_apl_2005,
Robertson_repproghys2006} who used screened exchange approximation to predict
vacancy energy levels including V$^-$ charge state. 
However, these calculations were performed using a small periodic supercell and 
therefore corresponded to extremely high vacancy concentrations. The quality
of the functional used is also largely unknown and needs independent 
verification. 

In this work we used much bigger supercells and a non-local functional to 
calculate optical excitation and thermal ionization energies of oxygen vacancies 
in five charge states. To relate these energies to experimental data we 
distinguish optical absorption/reflection type measurements involving Frank-
Condon type ({\it vertical}) excitations, and electrical thermal de-trapping 
measurements, where phonon-assisted electron excitations are accompanied by 
strong {\it lattice relaxation}. We focus on the results of so called de-trapping 
electical measurements \cite{Bersuker_micrrel_2004,Ribes_DevMatRel_2005,
Bersuker_irps_2006} 
which are interpreted in terms of thermal ionization of 
shallow electron traps and demonstrate that the interpretation 
is consistent with thermal ionization of negative V$^-$ and V$^{2-}$ vacancies. 

Our periodic non-local density functional calculations were carried out
using atomic basis set and a B3LYP hybrid density functional \cite{Becke} 
implemented in the CRYSTAL03 code \cite{Crystal_2003_manual}. This method 
reproduces band gaps of a range of transition metal oxides 
\cite{Cora_review_strbon2004} 
and allows us to carry out geometry optimization of defect structures, calculate 
defect electronic 
properties and excitation energies within the same method. We used all electron 
basis-set for oxygen atoms \cite{crystal_basis}, and an s,p,d valence basis set 
for Hf in conjunction with the relativistic small core effective potential due 
to Stevens {\it et al.} \cite{Hf_basis}. All calculations were carried out in 
the 96 atoms supercell of a monoclinic (m)-HfO$_2$ with a mesh of 36 k points in 
the irreducible 
Brillouin zone. The defect structures were optimized to the atomic forces below 
0.03 eV \AA $^{-1}$. The compensating uniform background potential method was 
used for charged defects calculations \cite{Leslie-Gillan,gavartin_jap2005}.   
 \begin{figure}
 \includegraphics[scale=0.45]{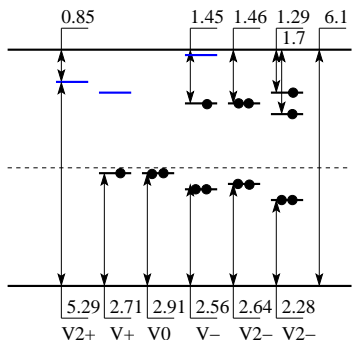}
 \caption{Defect energy diagram for the 4-coordinated oxygen vacancy in the 
 m-HfO$_2$.} 
 \label{fig_1}
 \end{figure}  
The positions of occupied and un-occupied single-electron defect levels in the 
gap of m-HfO$_2$ predicted by periodic B3LYP calculations are presented in 
Figure \ref{fig_1}. The results of the calculations can be summarized as 
follows: \\
1. The calculated single-electron band gap in m-HfO$_2$ is 6.1 eV. 
We note, that single-electron estimate neglects excitonic effects, and effects 
of phonon broadening which reduce measured optical band gap of HfO$_2$ by ~0.8 
eV \cite{Sayan_jap2004}. Taking this into account, predicted E$_g$ is in good 
agreement 
with the experimental optical gap of 5.6-5.9 eV \cite{afanasev04_offs}.  \\  
2. Oxygen vacancy in the monoclinic HfO$_2$ may exist in five charge states 
+2, +1, 0, -1, -2 corresponding to up to four extra electrons in the vicinity of 
the vacant O$^{2-}$ site. \\ 
3. The preferential site for the oxygen vacancy formation in m-HfO$_2$ depends 
on the charge state:  V$^{2+}$ and V$^+$ are more stable in the 3-fold 
coordinated 
sites, whereas the V$^0$, V$^-$, and V$^{2-}$ are energetically more favorable 
at the 4-fold coordinated sites (see also refs. 
\cite{Foster_hfo2_prb2002, nano_book_2003}). The difference in formation 
energies between 3- and 4-coordinated sites of negatively charged vacancies is 
about 0.2 eV. \\
4. The density of electrons in the +1 and zero charge vacancy 
is peaked at the vacant site with significant admixture of d-states of the 
nearest neighbor hafnium ions, similar to plane wave DFT calculations 
\cite{Foster_hfo2_prb2002,nano_book_2003}. The single-electron states of these 
electrons are located roughly in the middle of the gap (Fig. 1), significantly
lower than those reported by Robertson \cite{Xiong_Robertson_MicEng_2005,
Xiong_Robertson_apl_2005,Robertson_repproghys2006}\\
5. Both one and two extra electrons added to a neutral vacancy form more 
diffuse asymmetric states localized mainly on 3 (out of 4) Hf ions surrounding 
the vacancy. The ground state of the V$^{2-}$ is spin singlet, which is
 ~0.2 eV lower then the spin triplet configuration (cf. Fig. \ref{fig_1}).\\ 
6. The change of the vacancy charge state is accompanied by significant 
displacements of the nearest neighbor (NN) Hf atoms and next shell of oxygen 
atoms (NNN) (see also ref. \cite{Foster_hfo2_prb2002}.
The NN Hf ions displace from their perfect lattice position approximately symmetrically. 
This displacement is away from the V$^{2+}$ and V$^{+}$ 
by 11\% and 5\% of a typical Hf-O distance, but {\it towards} the V$^{-}$ and V$^{2-}$  
by 4\% and 8\% ,respectively. The displacements of the NNN oxygen ions are substantially 
smaller and directed away from the negative vacancies. \\
7. The character of electron density distribution, strong lattice relaxation and 
relatively shallow single-electron levels (Fig. \ref{fig_1}) suggest that 
trapping of extra electrons in negative V$^-$ and V$^{2-}$ vacancies is 
essentially polaronic in nature. The third and forth electrons induce strong 
lattice polarisation, which in turn, creates the potential well for these 
electrons. 
 \begin{figure}
 \includegraphics[scale=0.32]{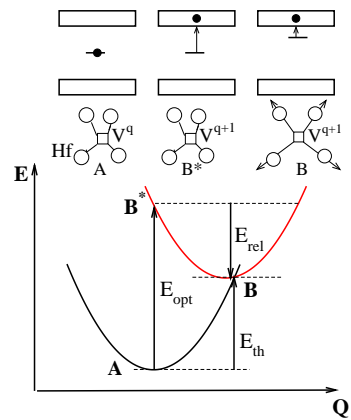}
 \caption{Schematic potential energy surface (PES) for optical and thermal 
 detrapping processes. Generalized displacement of the Hf atoms, Q, and energy
 scale is chosen to represent ionization of V$^{2-}$ vacancy. The band diagram 
 schematic illustrates optical and thermal ionization processes, respectively 
 without (B*) and with (B) lattice relaxation.}    
 \label{fig_2}
 \end{figure} 
Now we discuss a relation between the single-electron spectrum of Fig. 
\ref{fig_1} and experimental trap ionization measurements.
For optical type measurements, such as optical absorption, 
and reflection ellipsometry \cite{Takeuchi_jvst_2004}, using energy differences 
between single-electron levels can provide reasonable accuracy 
\cite{Foster_hfo2_prb2002,nano_book_2003}. The situation is different in 
electrical measurements, which include thermal excitation of trapped 
electrons into the conduction band of HfO$_2$ \cite{Bersuker_irps_2006}
(see inset of Fig. \ref{fig_2}). 

The difference between the optical and thermal 
ionization can be qualitatively explained by the potential energy surface 
diagram depicted in Fig. \ref{fig_2} and earlier discussed for other systems
(e.g. in ref.\cite{VandeWalle_jap2004}). Fast optical ionization of a trap of charge 
$q$ (state A) into the bottom of the conduction band (state B$^*$) corresponds to a 
Frank-Condon type transition, E$_{opt}$, when no lattice relaxation associated 
with the removed charge occurs. In contrast, a much slower 
thermal ionization of a trap, E$_{th}$, is a phonon assisted process, during which 
the system transfers into a fully relaxed trap state $q+1$ 
and the electron delocalized at the bottom of the conduction band (state B). 
The ionized trap induces lattice polarization, which in case of the oxygen 
vacancy is mainly associated with strong displacements of the NN Hf ions. 
This generalized relaxation coordinate is denoted as Q in the diagram in 
Fig. \ref{fig_2}. E$_{th}$ is approximately given by:
\begin{equation}
E_{th}=E_{opt}-E_{rel},  \label{eq_1}
\end{equation}
where E$_{rel}$ is the lattice relaxation energy. Optical and thermal ionization 
energies can be calculated using the 
{\it total} energies of systems in different charge states as follows:     
\begin{equation}
E_{opt}(V^{q})= E_{q}(V^{q+1})-E_{q}(V^{q})+E^{-}-E^{0} ; \label{eq_2}
\end{equation}
\begin{equation} 
E_{rel}(V^{q+1})= E_{q+1}(V^{q+1})-E_{q}(V^{q+1}). \label{eq_3}
\end{equation}                                               
The notations in Eqs. \ref{eq_2}-\ref{eq_3} have the following meaning:
E$^0$ - the total energy of the perfect HfO$_2$ crystal;
E$^-$ - the total energy of the perfect HfO$_2$ crystal with an electron at the 
bottom of the conduction band; E$_q$(V$^{q}$) - the total energy of HfO$_2$ with 
the 
vacancy in a charge state $q$ (q=+2,+1,0,-1,-2) in the optimized geometry;
E$_q$(V$^{q+1}$) - the total energy of the vacancy in the charge state $q+1$ but 
at the equilibrium geometry corresponding to the vacancy in the charge state 
$q$. We assume that the vacancy is well separated from the interface, and the 
optical and 
relaxation energies discussed here, depend only on the bulk properties of 
HfO$_2$ and not on the Si or metal band alignment. 
 \begin{table}
 \caption{Optical excitation energies, E$_{opt}$, relaxation energies,
 E$_{rel}$, and thermal activation 
 Energies, E$_{th}$, for oxygen vacancies in m-HfO$_2$ calculated according to 
 Eqs. \protect\ref{eq_1}-\protect\ref{eq_3}.} 
 \begin{tabular}{|l|rrr|}
 \hline
 q & E$_{opt}$ & E$_{rel}$ & E$_{th}$ \\
 \hline
 V$^+$ & 3.33 & 1.01 & 2.32 \\ 
 V$^0$ & 3.13 & 0.80 & 2.33 \\ 
 V$^-$ & 1.24  & 0.48 & 0.76 \\ 
 V$^{2-}$ & 0.99 & 0.43 & 0.56 \\ 
 \hline
 \end{tabular}
 \label{tab_1} 
 \end{table}
The calculated values of E$_{opt}$, E$_{rel}$ and E$_{th}$ are summarized in 
Table 
\ref{tab_1}. We note that optical ionization energies, E$_{opt}$ calculated as 
total energy differences in Eq. \ref{eq_2} are close to single-electron energy 
differences (Fig. \ref{fig_1}). 
This is to be expected, since the total energy differences for the systems with 
N and N$\pm$1 electrons in the DFT calculations are related to the single 
electron energies of the highest occupied and lowest unoccupied 
states\cite{Kantorovich_book}.  

The thermal ionization energies E$_{th}$ are, however, 0.5 to 1.0 eV {\it 
smaller} than the optical energies due to the large 
lattice relaxation associated with the change of the charge state of the 
vacancy. Although the single particle energies of V$^-$ and 
V$^{2-}$ are very similar (Fig. \ref{fig_1}), thermal ionization energies of 
these defects differ by ~0.2 eV. These values are consistent with 0.35, 
0.5 eV activation energies extracted from thermal de-trapping kinetics 
measurements 
\cite{Bersuker_irps_2006,Bersuker_micrrel_2004,Ribes_DevMatRel_2005}. 
On the other hand, very large 
de-trapping energies for the V$^+$ and V$^0$ vacancies rule out these species as 
possible shallow electron traps. We should note that accounting for the 
thermal broadening of the defect levels and band tails, which will be 
discussed elsewhere, would improve the agreement between 
the calculated thermal ionization energies and the experimental 
energies even further. 

To summarize, we have calculated the positions of single-electron levels for 
five charge states of oxygen vacancy in m-HfO$_2$ and related them to the 
existing experimental data. The results of trap discharging measurements 
\cite{Bersuker_irps_2006,Bersuker_micrrel_2004,Ribes_DevMatRel_2005} are consistent with thermal ionization 
of negatively charged V$^-$ and V$^{2-}$ oxygen vacancies. These results further 
support the common assumption that oxygen vacancies are likely candidates for 
intrinsic electron traps in these devices and suggest that negative oxygen 
vacancies can be responsible for $V_t$ instability.

%\bibliography{vacancy_hfo2}

\end{document}